\documentclass[twocolumn, epsf, prb, amsmath, amssymb, showpacs, superscriptaddress]{revtex4-1}
\usepackage{bm}
\usepackage{graphicx}
\usepackage{amsmath}
\usepackage{amssymb}
\usepackage{sidecap}
\usepackage{dcolumn}
\usepackage{color}
\usepackage{epstopdf}
\usepackage{ifpdf}
\usepackage{sidecap}
\usepackage{wrapfig}

\begin{document}
\title{Interplay between snake and quantum edge states in a graphene Hall bar with a pn-junction}

\author{S. P. Milovanovi\'{c}} \email{slavisa.milovanovic@uantwerpen.be}
\affiliation{Departement Fysica, Universiteit Antwerpen, \\
Groenenborgerlaan 171, B-2020 Antwerpen, Belgium}

\author{M.~Ramezani Masir}\email{mrmphys@gmail.com}
\affiliation{Departement Fysica, Universiteit Antwerpen, \\
Groenenborgerlaan 171, B-2020 Antwerpen, Belgium}
\affiliation{Department of Physics, University of Texas at Austin,\\
2515 Speedway, C1600 Austin, TX 78712-1192}

\author{F.~M.~Peeters}\email{francois.peeters@uantwerpen.be}
\affiliation{Departement Fysica, Universiteit Antwerpen, \\
Groenenborgerlaan 171, B-2020 Antwerpen, Belgium}

\begin{abstract}
The magneto-  and  Hall resistance of a locally gated cross shaped graphene Hall bar is calculated. The edge of the top gate is placed diagonally across the center of the Hall cross. Four-probe resistance is calculated using the Landauer-B\"{u}ttiker formalism, while the transmission coefficients are obtained using the non-equilibrium Green's function approach. The interplay between transport due to edge channels and snake states is investigated. When two edge channels are occupied we predict oscillations in the Hall and the bend resistance as function of the magnetic field which are a consequence of quantum interference between the occupied snake states.

\end{abstract}


\date{\today}

\maketitle
%\linespread{0.9}

Although graphene was introduced theoretically into the scientific community long time ago the great interest for this material started with its experimental realisation \cite{cgr}. High carrier mobility and a mean free path\cite{cmob1} that exceeds 1 $\mu$m together with a linear band structure\cite{cgp1, cgp2} are the first and foremost features that put this zero-gap semiconductor to the center of attention in electronic transport research. Ballistic transport allows the observation of transverse magnetic focusing\cite{cfoc1} as well as room temperature quantum Hall effect \cite{cgp2}.  

Graphene's chiral massless particles and a linear spectrum near the $K$ and $K'$ points cause perfect transmission through arbitrarily high and wide barriers\cite{ckt1, ckt2}. Recently, a graphene pn-junction was realized experimentally\cite{cf9, cf10} where separate control of carrier density in both regions could be obtained by using a pair of gates. The density in each region could be varied across the charge neutrality point, allowing pn-, pp-, and nn-junctions to be formed at the interface within a single graphene sheet. 

The presence of the pn-interface in graphene allows the formation of special propagating states along it - called \textit{snake states} \cite{css1, css2, cmy1, cmy3}. Due to the Lorentz force in combination with a change of sign of the carriers on the different sides of the pn-junction causes the bending of the current towards the pn-interface resulting in a channel of high mobility carriers along the pn-junction. In the case of multiterminal devices this can be a useful mechanism to control the output of the device\cite{cmy3}. The existence of snake states in graphene was predicted theoretically and attempts to observe them experimentally were undertaken recently.

\begin{figure}[htbp]
\begin{center}
\includegraphics[width=4.5cm]{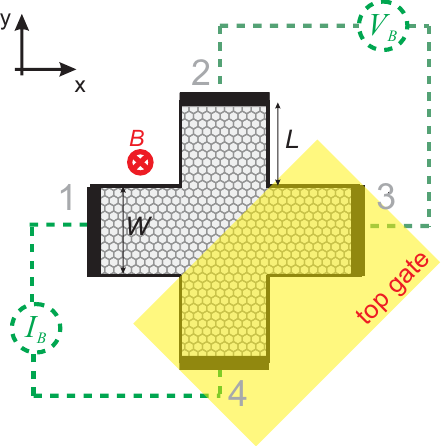}
\caption{(Color online) Schematics of a Hall bar structure with a tilted pn-junction.} \label{fstr}
\end{center}
\end{figure}

 The device investigated in this paper is schematically presented in Fig. \ref{fstr}. Numerical simulations of the electrical transport properties were performed using the Kwant code \cite{ckwant} for $W = L = 50$ nm. This python package uses non-equilibrium Green's functions (NEGF) to simulate transport through two-dimensional (2D) systems of arbitrary shape. Our system consists of a central, cross shaped scattering region, which is connected to four electrodes. In our case two terminals have zigzag edges (terminals 1 and 3) and the other two have armchair edges (terminals 2 and 4). Calculations are carried out for a fixed value of the Fermi energy, $E_F$, while the value of the applied perpendicular magnetic field  and the potential $V$ of the top gate are varied.
%
%Dirac fermions are described by the standard Dirac Hamiltonian:
%%
%\begin{equation}
%\label{edir}
%H = v_F\vec{\mathbf{\sigma}}(\vec{\mathbf{p}} + e\vec{\mathbf{A}}),
%\end{equation}
%%
%where $v_F$ is the Fermi velocity, $\vec{\mathbb{\sigma}} = (\sigma_x, \sigma_y)$ are the Pauli matrices, $\vec{\mathbf{p}}$ is the momentum, and $\vec{\mathbf{A}}$ is the magnetic vector potential. 
The magnetic field is introduced by replacing the hopping parameter $t$ in the tight-binding Hamiltonian\cite{ckwant}, using Peierls phase approximation\cite{cpeierls}, by $te^{i 2\pi \Phi_{ij}}$, where $\Phi_{ij} =e/h \int_{r_i}^{r_j} \vec{A} d\vec{r}$.  We choose the gauge  as
\begin{equation}\label{egau}
\vec{A} =  \frac{B}{2}\left[ x \sin\theta - y \cos\theta
\right]
\left(
\begin{array}{c}
\cos\theta \\
\sin\theta
\end{array}
\right),
\end{equation}
where $\theta$ is the angle between the $x$-axis and a specific lead. The gauge given by Eq. \eqref{egau} is used because it allows the transition between $\vec{A} = (-By,0)$ in leads parallel to the $x$-axis and the gauge $\vec{A} = (0,Bx)$ in the leads parallel to the $y$-axis. This is important for the proper construction of the leads and  their connection to the scattering region\cite{ckwant}. Transport properties are calculated using the Landauer-B\"{u}ttiker formalism \cite{cbut} which connects transmission probabilities obtained using the NEGF method with a measurable quantity - resistance. This method is widely used for multiterminal structures due to its simplicity and adaptability  to changes in the number of leads and their position. 

First, we will test our model for the case when the top gate is switched off, i.e. the applied potential is set to zero in the whole system. We apply a constant magnetic field, $B = 10$ T, while the electron density, $n_s$, is varied across the Dirac point.
\begin{figure}[htbp]
\begin{center}
\includegraphics[width=8cm]{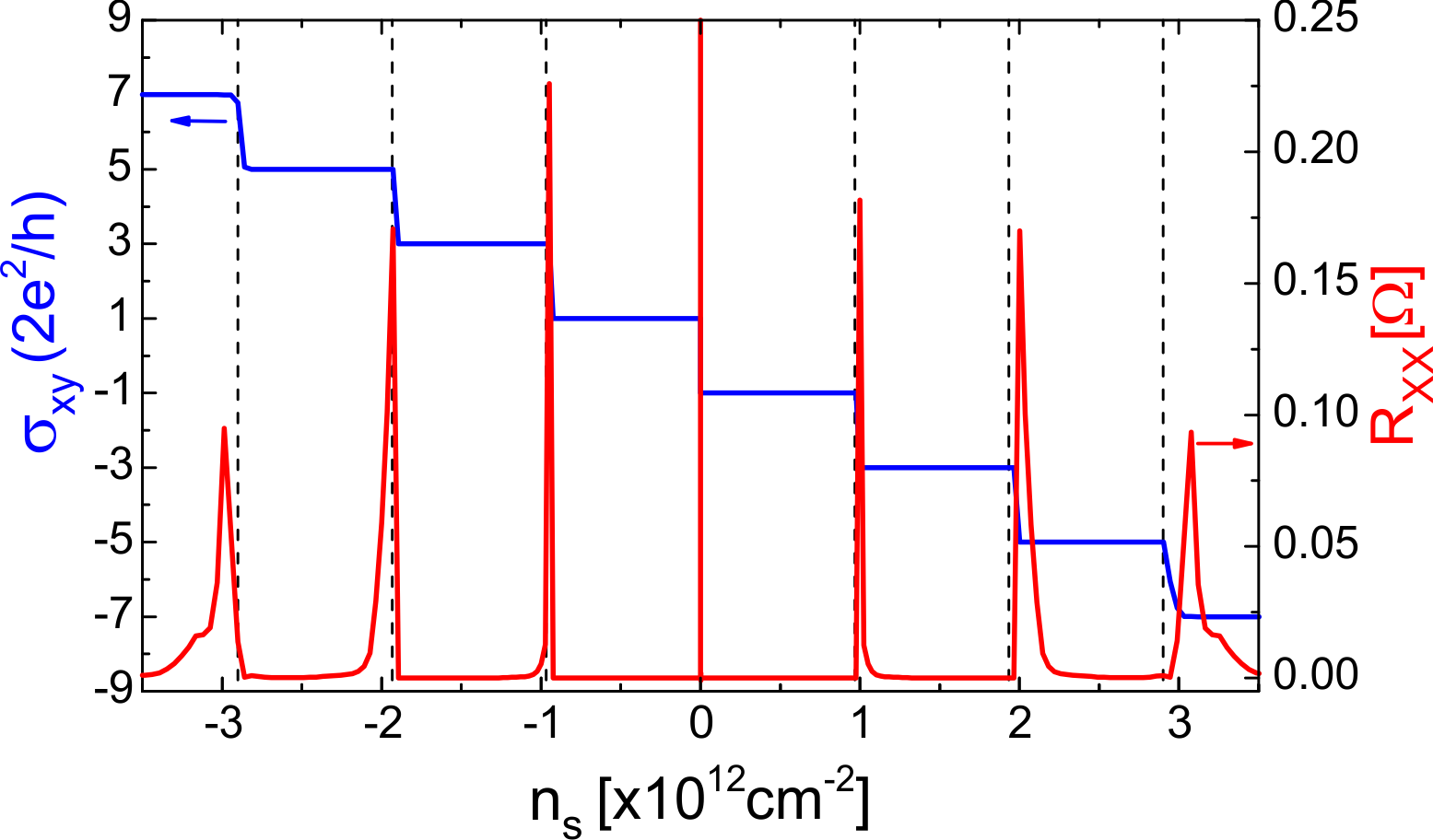}
\caption{(Color online) Hall conductance and longitudinal resistance for a Hall bar in a uniform magnetic field $B = 10$ T. Dashed lines show the position of the bulk Landau levels when moving through the Fermi level. Calculations are done for a Hall bar with $W = L = 50$ nm.} \label{fhbr1}
\end{center}
\end{figure}
\begin{figure*}[htbp]
\begin{center}
\includegraphics[width=16cm]{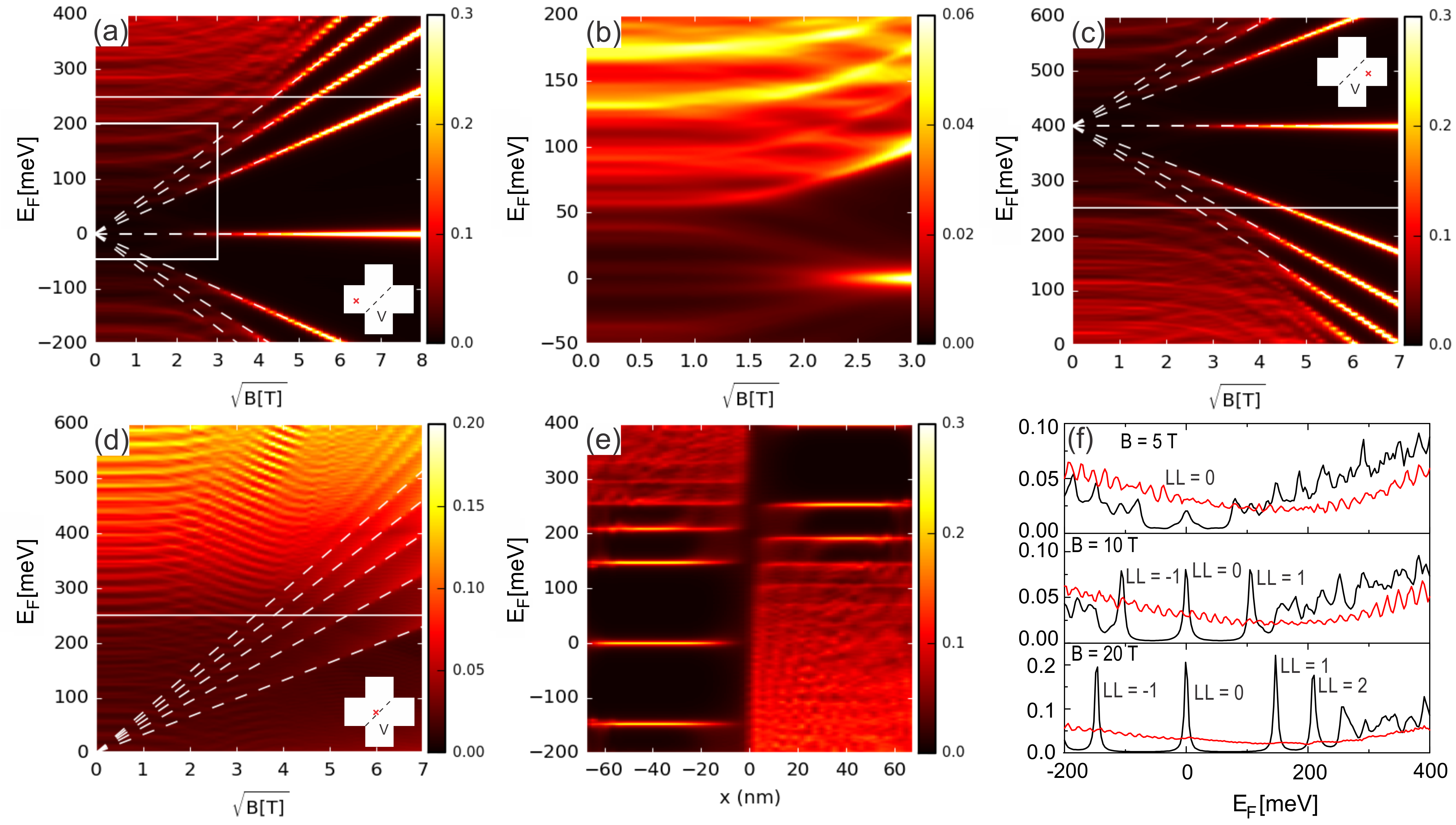}
\caption{(Color online) (a) LDOS in the  n-region of the pn-junction as indicated by the cross in the inset. (b) Zoom of LDOS from (a) shown by a white rectangle. (c) Same as (a) but for a point in the p-region. (d) Same as (a) but for a point that is very close to the pn-interface. (e) LDOS as a function of $x-$ coordinate and the Fermi energy and $B = 20$ T. (f) Cross-sections of LDOS from (a) and (d) given by black and red curves, respectively, for different values of the magnetic field.} \label{fldos1}
\end{center}
\end{figure*}
Results of the simulation are shown in Fig. \ref{fhbr1}. We see quantization steps of $4e^2/h$ in the Hall conductivity, $\sigma_{xy} = 1/R_{13,24}$, where the factor 4 comes from the spin and valley degeneracy, and the absence of a zero Hall plateau which is a hallmark of graphene. Each step in the conductance is followed by a peak in the longitudinal resistance, $R_{xx} = R_{13,13}$. Both events occur when a Landau level (LL) moves through the Fermi level. The position of the peaks agrees with the position of the Landau levels for bulk graphene shown in Fig. \ref{fhbr1} by the dashed lines except for the higher LLs where confinement effects start to play a role.

Next, we will examine how the response of our system is changed when the top gate is switched on, i.e. a potential step is added to the system, as presented by the yellow region in Fig. \ref{fstr}. This system configuration was recently investigated for a large, micron-size Hall bar \cite{cmy3} within a semiclassical model \cite{cbm}. Simulations revealed oscillations in the bend resistance, $R_B = R_{14,32}$, which were  linked with the existence of snake states around the pn-interface\cite{cmy3}. It was shown that the position of the peaks in the resistance could be accurately described by the following simple formula\cite{cmy3},
\begin{equation}\label{esc1}
\begin{array}{c}
\displaystyle{B_i^{peak} = \frac{2 E_F i}{ev_Fl_{pn}}[\eta + 1],~~~~i = 1,2,...,} \\
%\displaystyle{B_i^{dip} = 2d[i(1+\eta)-1],~~~~i = 1,2,..., }
\end{array}
\end{equation}
where $l_{pn}$ is the length of the pn-interface, $\eta = \left|(E_F-V)/E_F\right| $, $v_F$ is the Fermi velocity and $i$ is an integer. Logic behind this formula is that a peak in the resistance will appear if the length of the pn-interface is equal to an even integer multiple of the cyclotron radius. However, for small system sizes and/or high magnetic fields quantization effects are expected to make this formula inapplicable. 

Quantum effects are clearly visible in the local density of states (LDOS) plots of the system as shown in Fig. \ref{fldos1}, for an applied  potential $V = 400$ meV. Figs. \ref{fldos1}(a) and (c) show LDOS at points placed at different sides of the pn-junction (see cross symbol in inset of the figures). Plots show that for high fields the Landau levels (LLs) match the LLs of bulk graphene as given by the dashed lines. LLs on the side under the gate are shifted up by the value of the applied potential. White solid line at $E_F = 250$ meV shows the value of the Fermi energy for which the numerical simulations of the resistance will be performed later on. Although high, this $E_F$ was chosen such that the first three LLs are well developed and separated. Realistic computational times and memory requirements forced us to limit ourselves to nanosize systems. In our case, shown in Fig. \ref{fldos1}(b) we see that the zero and first LL start to develop at about $\sqrt{B} = 2.5 \sqrt{T}$. This can also be seen in Fig. \ref{fldos1}(f) where we selected a few cross-sections from Fig. \ref{fldos1}(a) (black curves) which show the formation of different LLs. Fig. \ref{fldos1}(d) shows the LDOS at a position that is closest to the pn-interface on the n-side and LLs are only distinguishable at high fields. In Fig. \ref{fldos1}(e) we show LDOS along the center of the Hall bar (pn-interface is at $x = 0$) for a magnetic field of $B = 20$ T.
\begin{figure*}[t]
\begin{center}
\includegraphics[width=16cm]{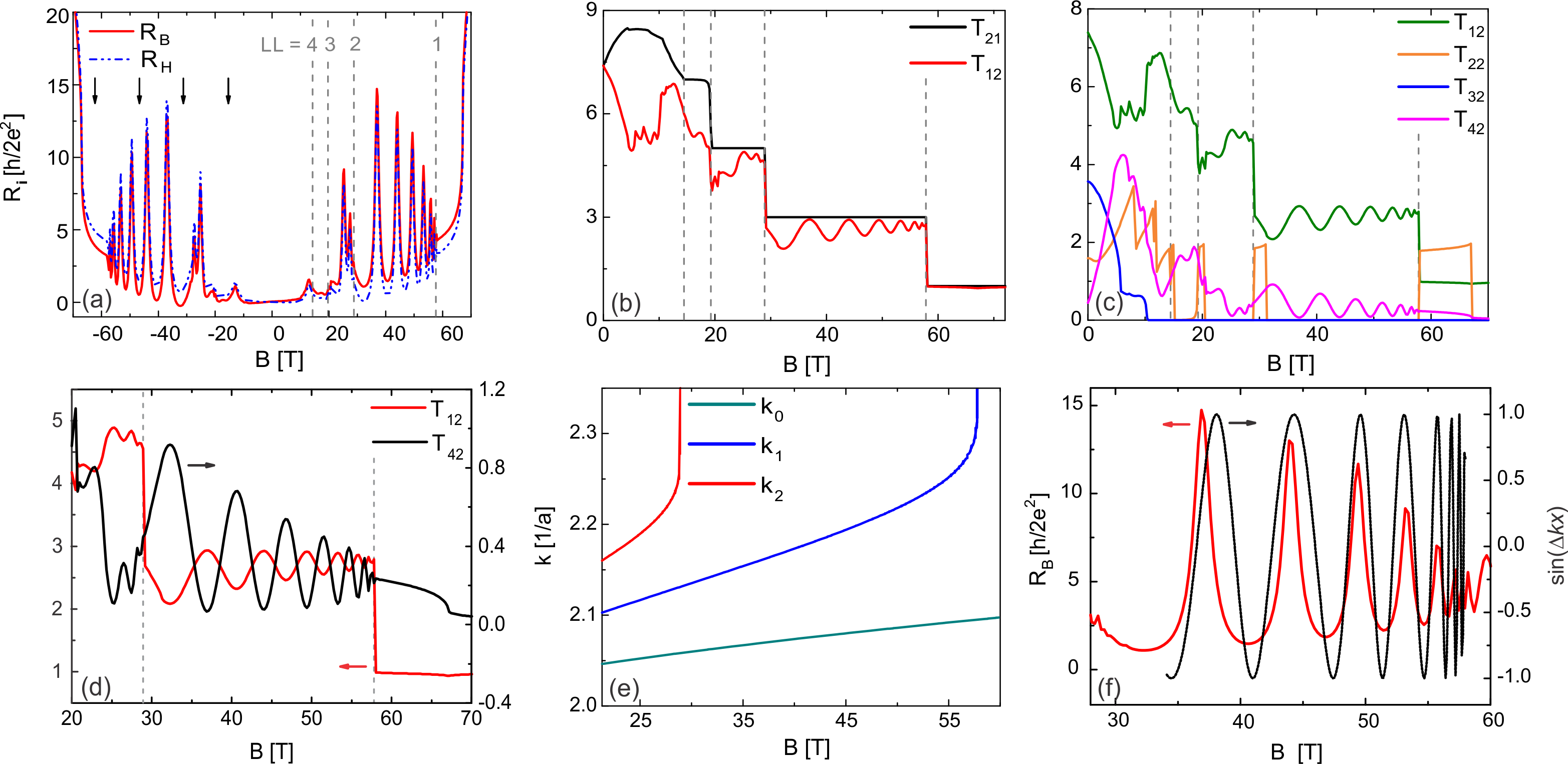}
\caption{(Color online) (a) Resistances $R_B$ and $R_H$ versus the applied magnetic field for $E_F = 250$ meV and $V = 400$ meV. The arrows indicate the position of the resistance peaks for $B<0$ due to the snake states as obtained from a semiclassical calculation. (b) Transmission coeficients $T_{21}$ and $T_{12}$. (c) Transmission coefficients in case of injection from lead 2. (d) Behavior of $T_{12}$ and $T_{42}$ in region between $LL=2$ and $LL=1$. (e) Fermi wave vector $k_n$ for occupied edge channel \textit{n} as a function of magnetic field. (f) $R_B$ and $\mathit{sin}(\Delta k x)$  for $\Delta k = k_1-k_0$ and $x = \sqrt{2}W$.} \label{fr1}
\end{center}
\end{figure*}

\begin{figure}[t]
\begin{center}
\includegraphics[width=7.5cm]{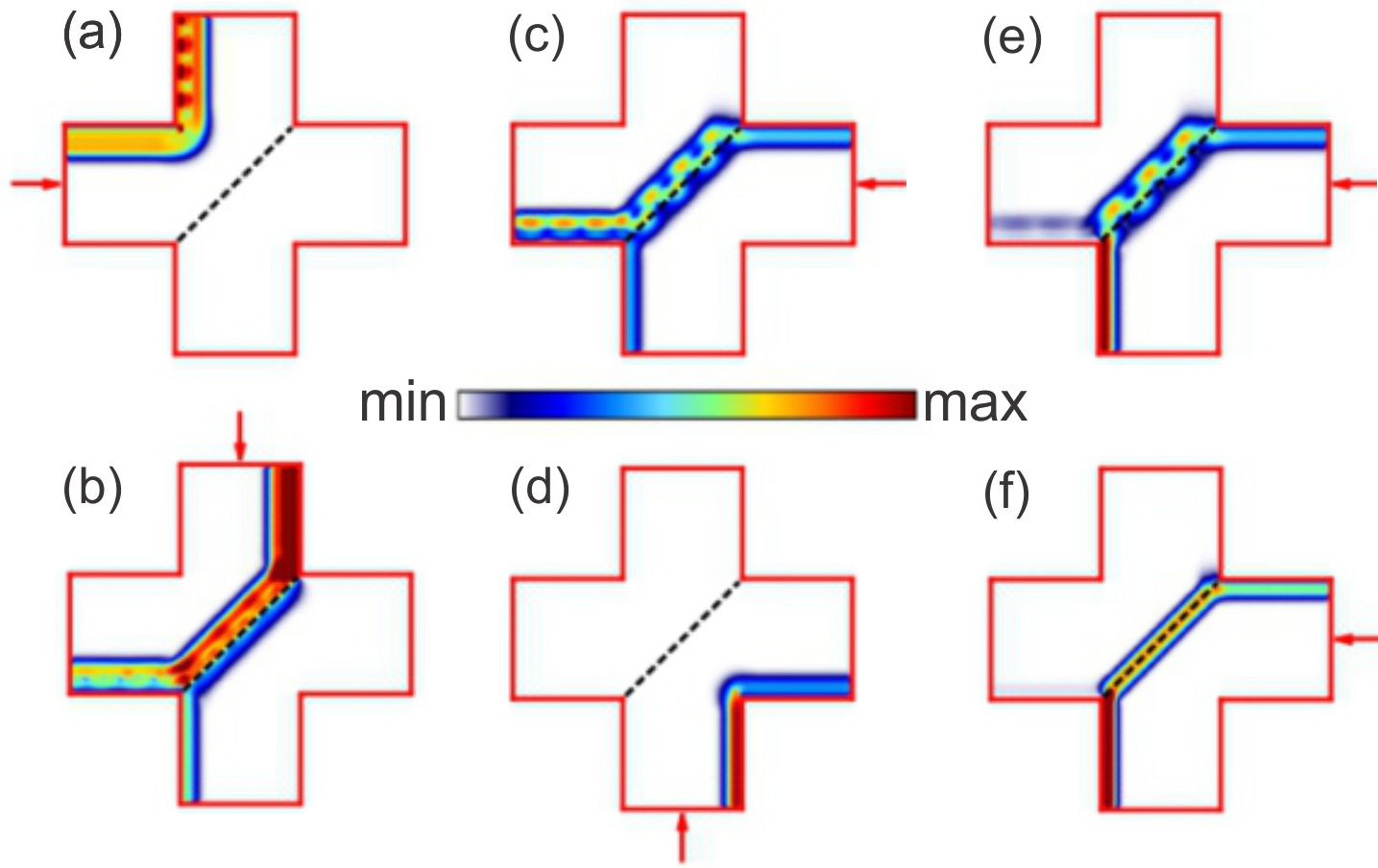}
\caption{(Color online) Current density for $E_F = 250$ meV, $V = 400$ meV and, (a)-(d) $B = 40$ T, (e) $B = 37$ T, (f) $B = 70$ T. Arrows show the injection lead and the black dashed line indicates the position of the pn-interface.} \label{fc1}
\end{center}
\end{figure}

Results for the bend resistance $R_B = R_{14,32}$ and the Hall resistance $R_H = R_{13, 24}$ are presented in Fig. \ref{fr1}(a) which show similar oscillating behavior and the absence of Hall plateaus. However, the position of these oscillations do not agree with the ones predicted by Eq. \eqref{esc1} which gives equidistant peaks in $B$ (as indicated by arrows in the negative $B$-range). Notice that beyond $LL = 2$ a series of peaks are found with increasing frequency and decreasing amplitude until $LL = 1$  when they disappear. To understand them we will make use of transmission and current density plots. Fig. \ref{fr1}(b) shows the transmission coefficients  $T_{21}$ and $T_{12}$ which exhibit quantization steps that match the LLs. However, $T_{12}$ exhibits much more interesting behavior with oscillations with maxima that approaches the value of $T_{21}$. To clarify the physics behind it we show the current density in Figs. \ref{fc1}(a, b). For positive magnetic field the current injected from lead 1 will follow the edges of the system and flow to lead 2. However, the current injected in lead 2 will reach the pn-interface and flow along it after which the injected beam splits between leads 1 and 4. The current flowing along the pn-interface is governed by snake states which collimate the injected beam. 
\begin{figure}[t]
\begin{center}
\includegraphics[width=8.5cm]{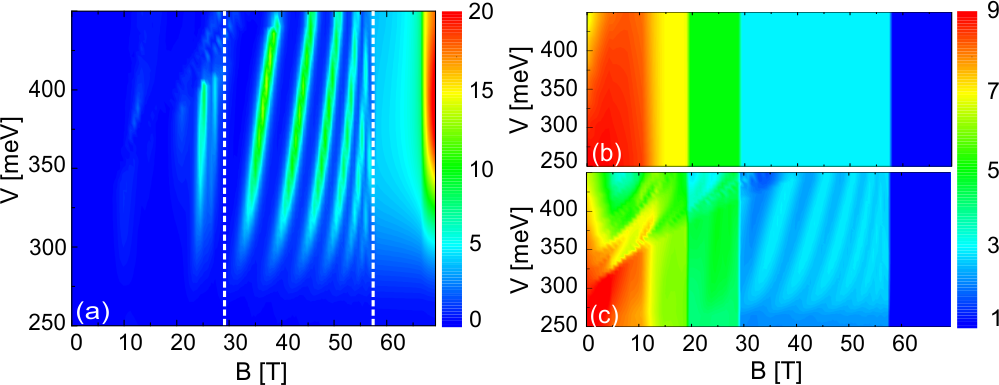}
\caption{(Color online) Contour plot of the (a) Bend resistance, $R_B$; the transmission coefficients (b) $T_{21}$ and (c) $T_{12}$ versus the magnetic field and the applied potential $V$. Values of resistance shown in the color bar are in units of $h/2e^2$. Calculations were preformed for $E_F = 250$ meV .} \label{fr3d1}
\end{center}
\end{figure}

Splitting of the injected beam is further investigated in Figs. \ref{fr1}(c) and (d). In Fig. \ref{fr1}(c) we present all transmission coefficients for injection from lead 2, while Fig. \ref{fr1}(d) shows the part of $T_{12}$ and $T_{42}$ between $LL=1$ and $LL=2$ (those LLs are indicated by vertical dashed lines). Fig. \ref{fr1}(c) tells us that the injected electron beam for $B > 10$ T (for which $2r_c<W$) is split into two parts - one that carries electrons to lead 1 and the other one that carries holes to lead 4. We can conclude this because the transmission coefficients $T_{12}$ and $T_{42}$ are much larger that the other two, which are practically zero, with a few exceptions when the LL is hit and an increase in reflection appears as shown by $T_{22}$. Notice that the transmission coefficients $T_{12}$ and $T_{42}$ are in antiphase. This is also seen in the current density plots shown in Figs. \ref{fc1}(c) and (e) where we plot the current injected from lead 3 in case when $T_{13}$ reaches its maximum (and $T_{43}$ is minimum) and the case when $T_{43}$ is maximum (and $T_{13}$ is minimum), respectively. These two plots show that the injected electron beam is well directed along the pn-interface and depending on the value of the cyclotron radius, $r_c$ (versus the length of the pn-interface), we can control in which lead it will end up. 

However, there is disagreement in the position of the peaks (indicated by the arrows in Fig. \ref{fr1}(a) for $B<0$) as predicted from the semiclassical results given by Eq. \eqref{esc1}. To address this question lets look at the behavior of $T_{12}$ and $T_{42}$ shown in Fig. \ref{fr1}(d). Oscillations that occur for magnetic fields between first and second LL resemble a beating signal which is reminiscent for the superposition of two sine waves with different frequencies. This is in fact what is happening. When we are between $LL=2$ and $LL=1$ two edge channels with wave vector $k_0$ and $k_1$ are occupied. The beating is a consequence of the superposition of these two edge channels which can be represented by two plane waves. A maximum appears whenever these two waves interfere constructively and we can write $sin(\Delta k \l_{pn}) = 2\pi m$, where $\Delta k = (k_1 - k_0)/2$. Fig. \ref{fr1}(e) shows variation of $k_n$ at the Fermi level with magnetic field, where $n$ is the channel number. As we approach the $n$-th LL, $k_n$ starts increasing rapidly which means that $\Delta k$ will also increase strongly and this is the reason for the decrease of the period of oscillations that we see in the resistance. In Fig. \ref{fr1}(f) we compare the resistance peaks with $\sin(\Delta k\l_{pn})$ for the case when only two edge channels are occupied. Notice the good agreement of the position of maxima and minima. The same reasoning can be applied for the case when three edge channels are occupied (the resistance peak between LL=3 and LL=2 in Fig. \ref{fr1}(a)). However, in this case the interference pattern can not be represented by a simple formula. When $B > 58$ T only one channel is occupied and the beating stops. Similar interference effect was found recently in Ref. \onlinecite{cnew} for edge channels in the case of transverse electron focusing of a normal 2D electron gas at the GaAs-AlGaAs interface.  

Figs. \ref{fr3d1}(a - c) show contour plots of $R_B$, $T_{21}$ and $T_{12}$ versus the magnetic field and the applied potential, $V$. Peaks in the resistance are clear and stable for all values of the applied potential. Notice that for $V = E_F = 250$ meV the carriers are depleted under the gate. $T_{21}$ exhibits clear quantization steps while  $T_{12}$ shows partial preservation of LLs, but more interestingly the interference pattern for the case when only two bands lay below $E_F$. Furthermore, the position of the peaks in $T_{12}$ agrees with the peaks in the resistance.

Finally, we studied the effect of disorder on the beats observed in the resistance. We examined cases of edge disorder as well as random vacancies. In the case of random vacancies we found that for a concentration of $0.02\%$ the beats have disappeared. However, in the case of edge disorder beats proved to be more robust surviving even for $3\%$ of atoms removed around the edges, i.e. very rough edges. 

In conclusion, in this letter we studied the electronic quantum response of a four-terminal graphene Hall bar. Dimensions of the structure were chosen in such a way that quantum mechanical effects are of significant importance. Simulations showed that the position of the peaks in the resistance are in disagreement with the ones predicted by classical simulations. Reason for this is the quantization of the cyclotron orbits. This can be best seen in the case when only two edge channels were occupied and oscillations in the resistance appear. We showed that these peaks are a consequence of the interference between snake states that were injected from edge channels. This was also confirmed by current density plots where we saw that the pn-interface collimates the injected electron beam and splits it in two parts which determines the output signal. Notice that peaks in the resistance were also predicted by our previous classical simulations but unlike there, in the present case the position of the peaks was determined by the interference of occupied edge channels.

This work was supported by the Flemish Science Foundation (FWO-Vl),
the European Science Foundation (ESF) under the EUROCORES Program
EuroGRAPHENE within the project CONGRAN and the Methusalem
Foundation of the Flemish government.

\end{document}